\documentclass[12pt]{article} 
\usepackage{amssymb}
\begin{document} 
\begin{center}
          {\large \bf Equations, chromopermittivity and instabilities\\
          of the quark-gluon medium } 

\vspace{0.5cm}                   
{\bf I.M. Dremin}

\vspace{0.5cm}              
          Lebedev Physical Institute, Moscow 119991, Russia

\end{center}
 
\begin{abstract}
The quark-gluon medium described by QCD equations is considered at high 
energies. Within the assumptions of the linear response theory the
chromopermittivity of the medium is introduced and it is argued that it exceeds 
1 at TeV energies. The dispersion equations show that the proper modes of the
medium reveal instability and the parton currents traversing it induce the 
emission of Cherenkov
gluons. Their distributions at LHC can differ from those typical for lower
energies of RHIC because they are determined by the high energy dependence
of the chromopermittivity while the latter ones arise due to collective 
resonance excitations.
\end{abstract}

\section{Introduction}

The quark-gluon medium formed in the collisions of high energy hadrons and
nuclei can possess some collective properties. At the classical level
QCD equations are similar to those of QED. Therefore it is quite natural to use 
the analogy with electrodynamical processes in matter. This approach is widely 
discussed (for recent review papers see, e.g., \cite{mrow, psmi}). The energy 
losses of charged particles in electrodynamics may be classified as 
elastic-collisional, radiative, Cherenkov, transition and wakes not to say about the 
ionization losses at low energies and nonlinear effects. The inelastic 
processes play especially important role in hadron collisions.

A deflection of particle trajectory from straight line 
is typical for the first two processes. Cherenkov and transition radiation 
as well as wakes behind the particle are described as proceeding at its
(almost) constant velocity. The particle charge induces polarization of
the medium (the permittivity different from 1). Namely this collective 
effect using the energy stored in the medium leads to Cherenkov 
radiation and wakes arising mostly from the matter
response induced by the polarization. We consider them here for
hadronic medium and, first of all, formulate the relevant equations.

The in-medium
equations of gluodynamics were proposed in \cite{inmed} (see also \cite{dpri}).
They differ from the in-vacuum equations by introducing a chromopermittivity
of the quark-gluon medium. Similar to the dielectric permittivity in
electrodynamics it describes the linear response of the matter to passing 
partons. At the classical level the equations are completely analogous to
electrodynamical ones with chromopermittivity just replacing the dielectric 
permittivity. However the energy behavior (dispersion) of these permittivities 
may be different because the collective quasi-particle excitations differ.
Therefore the dispersion equations and their predictions for spectra and
instabilities differ. As an example, we consider a specific model for high
energy dependence of the chromopermittivity inspired by experimental data on
hadronic elastic scattering amplitudes. It is shown that in this model the
quark-gluon medium is unstable and high energy partons moving in it 
emit Cherenkov gluons due to collective polarization effects. At very high 
energies their angular and energy distributions may differ from those of 
Cherenkov photons in electrodynamics. These conclusions can be verified at LHC.

\section{The in-medium equations and the Cherenkov gluons}

In this section we briefly remind the main results of the paper \cite{inmed}.

The classical in-vacuum Yang-Mills equations are
\begin{equation}
\label{f.1}
D_{\mu}F^{\mu \nu }=J^{\nu },
\end{equation}
\begin{equation}
\label{1}
F^{\mu \nu }=\partial ^{\mu }A^{\nu }-\partial ^{\nu }A^{\mu }-
ig[A^{\mu },A^{\nu }],
\end{equation}
where $A^{\mu}=A_a^{\mu}T_a; \; A_a (A_a^0\equiv \Phi_a, {\bf A}_a)$ are the 
gauge field (scalar and vector) potentials, the color matrices $T_a$ satisfy
the relation $[T_a, T_b]=if_{abc}T_c$, $\; D_{\mu }=\partial _{\mu }-ig[A_{\mu }, \cdot], \;\; 
J^{\nu }(\rho, {\bf j})$ is a classical source current, $\hbar=c=1$ and the 
metric tensor is $g^{\mu \nu }$=diag(+,--,--,--).

In the covariant gauge $\partial _{\mu }A^{\mu }=0$ they are written as 
\begin{equation}
\label{f.2}
\square A^{\mu }=J^{\mu }+ig[A_{\nu }, \partial ^{\nu }A^{\mu }+F^{\mu \nu }],
\end{equation}
where $\square $ is the d'Alembertian operator. The classical gluon 
field is given by the solution of the corresponding abelian problem.

The chromoelectric and chromomagnetic fields are
\begin{equation}
\label{2}
E^{\mu}=F^{\mu 0 },
\end{equation}
\begin{equation}
\label{3}
B^{\mu}=-\frac {1}{2}\epsilon ^{\mu ij}F^{ij},
\end{equation}
or, as functions of the gauge potentials in vector notation,
\begin{equation}
\label{4}
{\bf E}_a=-{\rm grad }\Phi  _a-\frac {\partial {\bf A}_a}{\partial t}+
gf_{abc}{\bf A}_b \Phi _c,
\end{equation}
\begin{equation}
\label{5}
{\bf B}_a={\rm curl }{\bf A}_a-\frac {1}{2}gf_{abc}[{\bf A}_b{\bf A}_c].
\end{equation}
The equations of motion (\ref{f.1}) in vector form are written as
\begin{equation}
\label{6}
{\rm div } {\bf E}_a -gf_{abc}{\bf A}_b {\bf E}_c = \rho _a,
\end{equation}
\begin{equation}
\label{7}
{\rm curl } {\bf B}_a-\frac {\partial {\bf E}_a}{\partial t} - gf_{abc}
(\Phi _b {\bf E}_c+[{\bf A}_b {\bf B}_c])= {\bf j}_a.
\end{equation}
                                                                  
Analogously to electrodynamics, the medium is accounted for if $\bf E$ is
replaced by ${\bf D} =\epsilon {\bf E}$ in $F^{\mu \nu} \;$, i.e. in Eq.
(\ref{2}). Therefore (\ref{6}), (\ref{7}) in vector form are
most suitable for generalization to the in-medium case.

In terms of potentials the equations of {\it in-medium} gluodynamics are cast 
in the form
\begin{eqnarray}
\bigtriangleup {\bf A}_a-\epsilon \frac{\partial ^2{\bf A}_a}{\partial t^2}=
-{\bf j}_a -
gf_{abc}(\frac {1}{2} {\rm curl } [{\bf A}_b, {\bf A}_c]+
\epsilon \frac {\partial }
{\partial t}({\bf A}_b\Phi _c)+\frac {1}{2}[{\bf A}_b {\rm curl } {\bf A}_c]-  \nonumber \\
\epsilon \Phi _b\frac 
{\partial {\bf A}_c}{\partial t}- 
\epsilon \Phi _b {\rm grad } \Phi _c-\frac {1}{2} gf_{cmn}
[{\bf A}_b[{\bf A}_m{\bf A}_n]]+g\epsilon f_{cmn}\Phi _b{\bf A}_m\Phi _n), 
\hfill \label{f.6}
\end{eqnarray}

\begin{eqnarray}
\bigtriangleup \Phi _a-\epsilon \frac {\partial ^2 \Phi _a}
{\partial t^2}=-\frac {\rho _a}{\epsilon }+ 
gf_{abc}(-2{\bf A}_c {\rm grad }\Phi _b+{\bf A}_b
\frac {\partial {\bf A}_c}{\partial t}-\epsilon 
\frac {\partial \Phi _b}{\partial t}
\Phi _c)+  \nonumber  \\
g^2 f_{amn} f_{nlb} {\bf A}_m {\bf A}_l \Phi _b. \hfill  \label{f.7}
\end{eqnarray}
If the terms with explicitly shown coupling constant $g$ are omitted, one gets
the set of abelian equations, which differ from electrodynamical equations
by the color index $a$ only. We omit it in what follows.
The most important property of the solutions of these equations is that
while the in-vacuum ($\epsilon = 1$) equations do not admit any radiation
processes, it happens for $\epsilon \neq 1$ that there are solutions of
these equations with non-zero Poynting vector even in the classical approach.
Here ${\bf j}_a$ is treated as an external current ascribed to partons  
moving fast relative to the other partons "at rest". It is proportional to $g$
as seen from Eq. (\ref{f.11}). Thus the terms with explicitly shown $g$ in
Eqs (\ref{f.6}), (\ref{f.7}) are of the order $g^3$. The higher order 
corrections may be calculated (preliminary results are published in \cite{okun}) 
but we postpone their consideration for further publications.
An impact of internal currents is phenomenologically taken into account by the
chromopermittivity $\epsilon $. Its spatio-temporal dependence may be included
similar to electrodynamics. It is well known in electrodynamics that the
magnetic permeability is automatically taken into account if the proper 
dependence of $\epsilon $ (in the Fourier representation) on the frequency 
($\omega $) and the wave vector
($\bf k$) is used. However it is premature now to discuss these complications.

For the current with velocity ${\bf v}$ along the $z$-axis:
\begin{equation}
\label{f.11}
{\bf j}({\bf r},t)={\bf v}\rho ({\bf r},t)=4\pi g{\bf v}\delta({\bf r}-{\bf v}t)
\end{equation}
the classical lowest order solution of in-medium gluodynamics can be cast in 
the form \cite{inmed, kruk}
\begin{equation}
\label{f.12}
\Phi ^{(1)}({\bf r},t)=\frac {2g}{\epsilon }\frac {\theta
(vt-z-r_{\perp }\sqrt {\epsilon v^2-1})}{\sqrt {(vt-z)^2-r_{\perp} ^2
(\epsilon v^2-1)}},
\end{equation}
and
\begin{equation}
{\bf A}^{(1)}({\bf r},t)=\epsilon {\bf v} \Phi ^{(1)}({\bf r},t),   \hfill  \label{f.13}
\end{equation}
where the superscript (1) indicates the solutions of order $g$, 
$r_{\perp }=\sqrt {x^2+y^2}$ is the cylindrical coordinate; $z$ is the
symmetry axis. 

This solution describes the emission of Cherenkov gluons at the typical angle
\begin{equation}
\label{f.10}
\cos \theta = \frac {1}{v\sqrt {\epsilon }}.
\end{equation}
It is constant for constant $\epsilon >1$.

First experimental data about ringlike (at constant $\theta $) emission of
particles in hadron collisions at high energies were published in \cite{apan}. They were interpreted
as an effect due to Cherenkov gluons in \cite{d1, d0}.

The expression for the intensity of the radiation is given by the Tamm-Frank
formula (up to Casimir operators)
\begin{equation}
\label{f.17}
\frac {dE}{dl}=4\pi \alpha_S\int \omega d\omega (1-\frac {1}{v^2\epsilon 
(\omega )})\theta (v^2\epsilon (\omega )-1).
\end{equation}
For absorbing media $\epsilon $ acquires the imaginary part. The sharp front 
edge of the shock wave (\ref{f.12}) is smoothed.  The angular distribution 
of Cherenkov radiation widens. The $\delta $-function at the angle 
(\ref{f.10}) is replaced by the Breit-Wigner shape \cite {gr, dklv} 
with maximum at the same angle (but $\vert \epsilon \vert $ in place of 
$\epsilon $) and the width proportional to the imaginary part. This has been
used for fits of RHIC data as explained in the next section.

\section{The chromopermittivity}

In in-medium electrodynamics the permittivity of real substances depends on
$\omega $. Moreover, it has an imaginary part determining the absorption.
E.g., ${\rm Re }\, \epsilon $ for water \cite{ja} is approximately constant
in the visible light region ($\sqrt {\epsilon }\approx 1.34$), increases at
low $\omega $ (up to $\epsilon \approx 80$)
and becomes smaller than 1 at high energies, tending to 1 asymptotically as
\begin{equation}
{\rm Re }\epsilon _{ED}=1-\frac {\omega _L^2}{\omega^2},         \label{eed}
\end{equation}
where $\omega _L$ is the Langmuir (plasma) frequency. The absorption 
(${\rm Im }\, \epsilon_{ED} $) is very small for 
visible light but dramatically increases in nearby regions both at low and 
high frequencies. Theoretically this behavior is ascribed to various collective
excitations in water relevant to its response to radiation with different 
frequencies. Among them the resonance excitations are quite prominent (see, 
e.g., \cite{fe}). Even in electrodynamics, the quantitative theory of this 
behavior is still lacking, however. Moreover the formula (\ref{eed}) is purely 
electrodynamical one and does not take into account hadronic processes at 
extremely high energies.

Then, what can we say about the chromopermittivity?

Up to now, the attempts to calculate the chromopermittivity from first principles 
are not very convincing. It can be obtained from the polarization operator.
The corresponding dispersion branches have been computed in the lowest order
perturbation theory \cite{kk, kl, we} and in the framework of the thermal field
theories \cite{bi, rrs, amy}. The results with an additional 
phenomenological {\it ad hoc} assumption about the role of resonances were used
in a simplified model of scalar fields \cite{ko} to show that the 
chromopermittivity can be larger than 1, which admits Cherenkov gluons. 
Extensive studies were performed in \cite{dpri}.

In view of this situation, we prefer to use the general formulae of the 
scattering theory \cite{go} to estimate the chromopermittivity. It is related 
to the refractive index $n$ of the medium:
\begin{equation}
\label{f.18}
\epsilon =n^2
\end{equation}
and is expressed \cite{ja, go, scad} through the real 
part of the forward scattering amplitude ${\rm Re}F_0(\omega)$ of the refracted 
quanta\footnote{In electrodynamics these quanta are photons, in QCD they are 
gluons.} 
\begin{equation}
\label{f.19}
{\rm Re} \Delta \epsilon = {\rm Re} \epsilon (\omega ) -1= \frac {4\pi N_s 
{\rm Re} F_0(\omega)}{\omega ^2}=\frac {N_s\sigma (\omega )\rho (\omega )}{\omega }   
\end{equation}
with
\begin{equation}
{\rm Im} F_0(\omega )=\frac {\omega }{4\pi }\sigma (\omega ).
\end{equation}
Here $\omega$ denotes the energy, $N_s $ is the density of scattering centers,
$\sigma (\omega)$ the cross section and $\rho (\omega)$ the ratio of real to 
imaginary parts of the forward scattering amplitude $F_0(\omega)$. Thus the 
emission of Cherenkov gluons is possible only for processes with positive 
${\rm Re} F_0(\omega)$ or $\rho (\omega)$. Unfortunately, we are unable to 
calculate directly in QCD these characteristics of gluons and have to rely
on analogies and our knowledge of the properties of hadron interactions. 
The experimental facts we get for this medium are brought about only 
by particles registered at the final 
stage. They have some features in common, which (one may hope!) are also
relevant for gluons as the carriers of the strong forces. Those are the resonant
behavior of amplitudes at rather low energies and the positive real part of the
forward scattering amplitudes at very high energies for hadron-hadron 
($pp, Kp, \pi p$) and photon-hadron ($\gamma p$) processes as measured from the 
interference of the Coulomb and hadronic parts of the amplitudes. 
This shows that the necessary condition for Cherenkov effects might be
satisfied at least within these two energy intervals. This fact was used
to describe experimental observations at SPS, RHIC and cosmic ray energies. 

The first region is typical for the comparatively low energies of secondary
particles registered in SPS and RHIC experiments. ${\rm Re} F_0(\omega)$ is 
always positive (i.e., $\epsilon >1$) within the low-mass wings of the 
Breit-Wigner resonances. Therefore, Cherenkov gluons can be emitted in
these energy intervals. 

The asymmetry of the $\rho $-meson mass shape observed in leptonic decays 
of $\rho $-mesons created in nuclei collisions at SPS \cite{da} was explained by 
emission of low-energy Cherenkov gluons \cite{dnec, drem1}
inside the left (low mass) wing of the Breit-Wigner resonance. It is predicted
that this feature should be common for all resonances traversing the nuclear
medium. Some preliminary experimental indications which favor this conclusion 
have appeared for other resonances as well \cite{4, 5, Muto, 6, 7}.

The experimental data of STAR and PHENIX collaborations at RHIC 
\cite{fw, ph, ul, ajit, abel} on two- and three-particle azimuthal correlations 
also deal with rather low energies of secondary particles. 
The ringlike distribution of the particles around the (away-side) jet traversing
the quark-gluon medium was observed in the form of two humps when projected
on the diameter of the ring. That is completely analogous to what was done
by Cherenkov in his original publications (see, e.g., \cite{jel}).

Similar to electrodynamics \cite{gr}, it is possible to get the 
energy-angular spectrum of emitted gluons \cite{wake} per the unit length
\begin{equation}
\frac {dN^{(1)}}{d\Omega d\omega }=\frac {\alpha _SC}{2\pi }\left [
\frac {(1-x)\Gamma _t}{(x-x_0)^2+(\Gamma _t)^2/4}+\frac {\Gamma _l}{x}\right ], 
\label{9}
\end{equation}
where 
\begin{equation}
x=\cos ^2\theta, \;\;\;  
x_0=\epsilon_{1t}/\vert \epsilon _t \vert ^2v^2, \;\;\;                         
\Gamma _j=2\epsilon_{2j}/\vert \epsilon _j \vert ^2v^2, \;\;\;
\epsilon _j=\epsilon _{1j}+i\epsilon _{2j}.
\end{equation}
The first term in the brackets corresponds to the transverse gluon Cherenkov
radiation (index $t$ at $\epsilon _t$) and the second term to the radiation
due to the longitudinal wake (index $l$ at $\epsilon _l$). The transverse and
longitudinal components of the chromopermittivity  tensor are explicitly
indicated here even though they are equal in any homogeneous medium.
The real ($\epsilon_1$) and imaginary ($\epsilon_2$) parts of $\epsilon $ are 
taken into account. The angle $\theta $ is the polar angle if the away-side 
jet axis would be chosen as $z$-axis. The ringlike structure around this axis 
is clearly exhibited in (\ref{9}). The angle $\theta $ is related to the
laboratory polar ($\theta _L$) and azimuthal ($\phi _L$) angles in RHIC 
experiments as
\begin{equation}
x=\cos ^2 \theta = \sin ^2 \theta _L \cos ^2\phi _L . \label{tpl}
\end{equation}
Integrating (\ref{9}) over $\theta _L$ one gets \cite{dklv} the final formula 
to compare with the two-hump structure of azimuthal ($\phi _L$) correlations 
observed at RHIC. It is quite lengthy and not reproduced here. It is seen 
already from (\ref{9}) and (\ref{tpl}) that this projection of the
two-dimensional ring on its diameter is symmetrical about $\phi _L=\pi $ and 
exhibits humps whose positions are mostly determined by $\epsilon _1$ and 
widths determined by $\epsilon _2$. The first term in (\ref{9}) clearly
demonstrates the a'la Breit-Wigner angular hump which replaces the 
$\delta $-functional angular dependence characteristic for real $\epsilon $. 
For two-particle correlations in central Au-Au collisions at 200 GeV measured 
by STAR \cite{fw} and PHENIX \cite{ph} 
it was found in \cite{dklv} that $\epsilon_1$ and $\epsilon_2$ are ranging, 
correspondingly, from 5.4 to 9 and from 0.7 to 2. The results differed because 
of disagreement in peaks positions and widths in the old data of these 
collaborations with peaks at $\pi \pm 1.04$ and $\pi \pm 1.27$, correspondingly. 
The new data \cite{holz} agree that maxima are positioned at $\pi \pm 1.1$ so 
that $\epsilon _1\approx 6$ and $\epsilon _2\approx 1$ would be good estimates
\footnote{The constant values (no dispersion) are considered because the range
of pion energies within the ring is comparatively small.}.
The main conclusion about the large values of $\epsilon _1$ indicating on the 
non-gaseous matter (large $N_s$ in (\ref{f.19})) is supported in any case. For
three-particle correlations the peak is shifted to larger angles in the earlier 
data \cite{abel} and agrees with above estimates in the new data \cite{holz}.
The in-plane and out-of-plane structures of two- and three-particle correlations
in semi-central events \cite{holz} can be described in the same way. More
important is the fact that the wake contribution described by the second term 
in (\ref{9}) leads to the shift of the maxima \cite{wake} observed in the data 
\cite{holz} obtained with triggers positioned in-between (at $\pi /4$).

In principle, the two-hump structure may arise not only due to Cherenkov gluons 
but also as the Mach cone. However, there exists the clear distinctive
feature which favors the QCD interpretation compared with relativistic 
hydrodynamics. This is the property of the corresponding wakes.
The hydrodynamical Mach cone interpretation was not yet used 
for quantitative fits of the data because the particle yield from the wake 
\footnote{It is transverse in hydrodynamics (see \cite{mus1, mus2, rmu}).} 
moving opposite to the trigger jet largely overwhelms the weak Mach cone signal 
\cite{betz} for relativistic particles. It should produce a jet of particles 
in this direction (the strong away-side jet) that is not observed in experiment.

\section{Instabilities in the high energy region}

The specifics of RHIC experiments is that triggers register trigger 
jets with energies of several GeV emitted at large angles (close to $\pi/2$) 
to the collision axis.
In its turn, cosmic ray data \cite{apan} at energies corresponding to LHC 
energies require that beside effects induced by the comparatively low energy
(several GeV) partons there should be very high-energy Cherenkov gluons emitted by the 
ultrarelativistic partons moving along the collision axis as was first proposed
in \cite{d1, d0}. Let us note the important difference from pure electrodynamics, 
where $\epsilon <1$ at high frequencies (see (\ref{eed})). In hadron physics, 
${\rm Re} F_0(\omega)$ is positive (and consequently $\epsilon >1$
according to (\ref{f.19})) above some threshold as measured in experiment. 
The dispersion relations for particle scattering amplitudes explain it as 
the corollary of the cross sections 
increase at high energy\footnote{Surely, taking into account the hadronic 
channels of photon interactions at high energies one would come to the similar 
conclusions and the formula (\ref{eed}) will change.}. This could be of crucial
importance for experiments at the LHC opening a new channel for the medium
polarization effects.

In what follows, we concentrate on the high energy region. The analogy of the
two colliding nuclei considered as clouds of partons to the two colliding 
bunches of plasma (see, e.g., \cite{rukh}) will be used. However, the
permittivities differ in these cases. It leads to different conclusions.

The energy dependence of the factors in (\ref{f.19}) is not precisely known.
The total cross section increases but not faster than $\ln^2\omega $, the 
density of scatterers must saturate and the real part of the forward 
scattering amplitude is positive and, probably, decreasing. The increase of
total cross sections leads to positive ${\rm Re}F_0(\omega )$ according to
the dispersion relations \cite{dnaz}. Therefore, we will study the
properties of the quark-gluon medium in the model with chromopermittivity
behaving above some threshold as 
\begin{equation}
{\rm Re}\epsilon =1+\frac {\omega _0^2}{\omega^2},         \label{qcd}
\end{equation}
where $\omega_0$ is some real free parameter. In principle, it may depend on 
energy but at the beginning we consider it to be a constant. This is the 
simplest formula for the even (as required by the Kramers-Kronig relations) 
function of $\omega $ exceeding 1 and tending to 1 at
$\omega \rightarrow \infty $ satisfying the general analytical requirements
imposed on $\epsilon $. It reduces to 
electrodynamics with imaginary $\omega_0$ and is somewhat different from 
treatment in \cite{d1, d0} where $\omega _0^2=a\omega $ with constant $a$ was 
used\footnote{This contradicts to the requirement for ${\rm Re}\epsilon $ to be the
even function of $\omega $ but can be considered as a purely phenomenological 
fit in some restricted region of $\omega $.}.
In hadronic processes the threshold of positiveness of ${\rm Re}F_0(\omega )$
is near 100 GeV in the rest system of the target proton in the above mentioned
processes. If the experimental values of $\sigma (\omega )\rho (\omega )/\omega $
are plotted, they show the increase at the threshold, the maximum slightly 
below 1 TeV in the rest system and the subsequent decrease. Thus (\ref{qcd}) is
also motivated by Eq. (\ref{f.19}). It is premature now
for our purposes to consider more elaborate forms of the relation (\ref{qcd}).

For the dispersion law (\ref{qcd}) the classical (no recoil) coherence length is
\begin{equation}
l_{coh}= \frac {2\gamma ^2}{\omega }\left(1-\frac {\omega _0^2}{\omega ^2}
\gamma ^2+\theta ^2\gamma ^2\right)^{-1},    \label{lcoh}
\end{equation}
where $\gamma $ is the $\gamma $-factor of the colliding hadrons. This length
\footnote{Sometimes it is called as the formation length while the term
coherence length is ascribed to the factor $2\gamma ^2/\omega $ in front of the 
brackets in (\ref{lcoh}).}
is larger than for negative $\omega _0^2 $ and becomes especially
large at Cherenkov angles quadratically increasing with energy. The requirement
of positive $l_{coh}$ restricts the transverse momenta of emitted gluons
\begin{equation}
k_t^2 \geq \omega _0^2-\frac {\omega ^2}{\gamma ^2} \geq \omega _0^2-x^2m^2 
\geq 0,         \label{kt}
\end{equation} 
where $m$ is the proton mass and $x$ is the share of its momentum carried by 
the parton. For these inequalities to be valid at any $x$, one needs
$\omega _0\geq m\approx 1$ GeV.

Let us mention that the relation of the current with the field is determined
by the conductivity $\sigma _{i, j}(\omega , {\bf k})$  (in the tensor 
notation) as
\begin{equation}
{\bf j}_i(\omega , {\bf k})=\sigma _{i, j}(\omega , {\bf k}){\bf E_j}(\omega , {\bf k}).
\end{equation}
The permittivity is related to the conductivity as
\begin{equation}
\epsilon _{i, j}=1 +\frac {4\pi i}{\omega }\sigma _{i, j} . \label{cond}
\end{equation}
Thus, the positive sign in (\ref{qcd}) implies the negative inductive
conductivity, i.e. the damped induced current opposite to the direction
of the field. In electrodynamics, the negative conductivity arises due
to the presence of magnetic fields \cite{ryzh}. It is actively studied now
\cite{eles}.
Whether the dispersion law (\ref{qcd}) in the case of chromodynamics is 
related to internal chromomagnetic fields must be studied.  

The dielectric permittivity of the macroscopic matter (e.g., as given by 
(\ref{eed})) is usually considered in its rest system which is well defined. 
For collisions of two nuclei (or hadrons) such a system 
requires special definition (see \cite{inmed}). In particular, for fast 
forward moving partons the spectators (the medium) are formed by the partons 
of another (target) nucleus at rest. Thus we consider a problem of the 
system of the quark-gluon
medium impinged by a bunch of fast partons similar to the problem
of plasma physics, namely, that of the interaction of the electron
bunch with plasma \cite{aruk}. This differs from RHIC conditions when
partons scattered at $\pi /2$ were considered with the rest system of the 
medium coinciding with the nuclei center of mass system.

Two complications are to be considered.
First, there are selected directions of the current ($z$-axis) and its radiation.
This is cured by introducing the permittivity tensor as
\begin{equation}
D_i(\omega , {\bf k})=\epsilon _{ij}(\omega , {\bf k})E_j(\omega , {\bf k}).
\end{equation}
Second, the impinging bunch content is similar to that of the target 
(plasma-plasma collisions!) and its permittivity must also be accounted for. 
In the projectile system it is the same as the permittivity of the target in 
its rest system, i.e. given by (\ref{qcd}). Then one takes into account that
the total induced current is the sum of currents induced in the target and in
the projectile and Lorentz-transforms the projectile internal fields and 
currents to the target rest system where we 
consider the whole process as it is done in \cite{rukh, aruk, sruk, ginz}. Using 
the current conservation and classical field equations (\ref{6}), (\ref{7}) 
with $g=0$ one gets for non-zero components of the chromopermittivity tensor
\begin{eqnarray}
\epsilon_{xx}=\epsilon_{yy}=1+\frac {\omega_0^2}{\omega ^2}(1+\frac {1}{\gamma}),
\nonumber \\
\epsilon _{xz}=-\epsilon_{zx}=\frac {\omega_0^2k_t}{\omega ^2(\omega -k_z)\gamma },
\nonumber \\
\epsilon _{zz}=1+\frac {\omega_0^2}{\omega ^2}\left(1+\frac {k_t^2}{(\omega -k_z)^2
\gamma }\right ).                    \label{tens}
\end{eqnarray}
Here $k_t$ and $k_z$ are the transverse and longitudinal components of 
${\bf k}$. We use the approximation of high energies (large $\gamma \gg 1 $ 
factor, i.e. $v\approx 1$). The terms depending on $\gamma $ are due to the 
impinging partons. They can be omitted everywhere except the terms which 
determine the Cherenkov gluon radiation at $\omega -k_z\approx 0$.

The classical equations derived from (\ref{f.6}), (\ref{f.7}) and written
in the momentum space have solution if the following dispersion equation
is valid
\begin{equation}
{\rm det}(\omega, {\bf k})=\vert k^2\delta _{ij}-k_ik_j-\omega ^2\epsilon _{ij}
\vert =0.  \label{disp}
\end{equation}
It is of the sixth order in momenta dimension. However, the sixth order terms 
cancel and (\ref{disp}) leads to two equations (of the second order):
\begin{equation}
k^2-\omega ^2-\omega _0^2=0,     \label{plas}
\end{equation}
\begin{equation}
(k^2-\omega ^2-\omega _0^2)(1+\frac {\omega_0^2}{\omega^2})-
\frac {\omega _0^4k_t^2}{\omega ^2(\omega-k_z)^2\gamma }=0.  \label{bunch}
\end{equation}
They determine the internal modes of the medium and the bunch propagation
through the medium, correspondingly.

The equation (\ref{plas}) shows that the quark-gluon medium is unstable because
there exists the branch with ${\rm Im}\omega >0$ for modes $k^2<\omega _0^2$.
Thus the energy increase of the total cross sections is related to the
instability of the quark-gluon medium by the positiveness of 
${\rm Re}F_0(\omega )$ at high energies.

The equation (\ref{bunch}) has solutions corresponding to Cherenkov gluons 
emitted by the impinging bunch and determined by the last term in (\ref{bunch}).
They can be found at $\omega=k_z+\delta \; (\delta \ll \omega)$. For
$k_t=\omega _0$ one gets the solutions with
\begin{equation}
{\rm Im} \delta_1=\frac {\omega _0^2}{2k_z[2\gamma (1+\omega_0^2/k_z^2)]^{1/3}}.
\label{d1}
\end{equation}
For $k_t\neq \omega_0$ there is the solution with 
\begin{equation}
{\rm Im} \delta_2=\frac {\omega _0^2k_t}{k_z[\gamma \vert k_t^2-\omega _0^2\vert
(1+\omega_0^2/k_z^2)]^{1/2}}.                 \label{d2}
\end{equation}
It is well known (see \cite{kruk}) that the solutions of the disperion 
equation (\ref{disp}) determine the Green function of the system equations
\begin{equation}
G(t,z)=\frac {1}{2\pi^2}\int _{-\infty}^{\infty}dk\int _{C(\omega )}\frac {1}
{{\rm det}(\omega, {\bf k})}\exp (-i\omega t+ikz)d\omega,      \label{green}
\end{equation}
where the contour $C(\omega )$ passes above all singularities in the integral.
Therefore, the positive ${\rm Im}\delta _i $ in (\ref{d1}) and (\ref{d2})
correspond to the absolute instability of the system. Let us note that the
instability at $k_t=\omega _0$ is stronger than at $k_t\neq \omega _0$ 
approximately by the factor $\gamma ^{1/6}$ (this factor is about 4 times 
larger at LHC compared to RHIC). The instability exponent (\ref{d1}) decreases as 
$\gamma ^{-1/3}$ and is about 16 times smaller at LHC compared to RHIC.
It tends to zero asymptotically.

Thus Cherenkov gluons are emitted with constant transverse momentum 
$k_t=\omega _0$ and their number is proportional to 
$d\omega /(\omega )^2\theta (\omega -\omega _{thr})$ 
for $\epsilon (\omega )$ given by Eq. (\ref{qcd}) with account of the
threshold above which this equation is applicable. It differs from the 
traditional folklore of constant emission angle of Cherenkov radiation and the 
number of gluons $\propto d\omega $ (or the total energy loss proportional
to $\omega d\omega $). This difference is easily explained by Eqs
(\ref{f.10}), (\ref{f.17}) which give $\cos \theta $=const and the energy loss
$\omega d\omega $ for $\epsilon $=const and $k_t\approx \omega _0$ 
and $d\omega /\omega $ energy loss for $\epsilon = 1+\omega _0^2/\omega ^2$. 
The electrodynamical formulae for transition radiation (see, e.g., \cite{ryaz})
also get the singularity in the spectra at $k_t\approx \omega _0$
for such $\epsilon $. 

Therefore, the conclusions strongly depend on the realistic shape of
$\epsilon (\omega )$ (and according to (\ref{f.18}) on the refraction index $n$)
at high energies. It was shown \cite{d1, d0} that according
to experimental data on $pp$-scattering and eq. (\ref{f.19}) $\Delta \epsilon
(\omega )$ increases above the threshold, has a maximum about $ 10^{-5}$
at $\omega \approx 1$ TeV and then decreases. One easily gets this rough
estimate if inserts in Eq. (\ref{f.19}) the total $pp$ cross section about
60 mb, $\rho _{pp}\approx 0.1$ at $\omega \approx 1$ TeV and 
$N_s \approx 3m_{\pi }^3/4\pi $ (considering the proton as a single scattering
center). Then $\omega _0\approx 3$ GeV. From inequalities (\ref{kt}) it follows
that the coherence length is positive for $x<\omega _0/m\approx 3$, i.e. for all 
partons inside the impinging nucleon (nucleus). These inequalities must be 
valid for all gluons radiated with extremely high energies. If optimistic, 
one might expect 
some excess at transverse momenta from 1 to 10 GeV due to these processes. 
This intuitive estimate tries to account for variation of the above parameters
and possible role of the imaginary part of $\epsilon $.

The rate of the decrease of $\Delta \epsilon(\omega )$ is not
well determined (in \cite{d1, d0} it was used as $\omega ^{-1}$). Thus the 
step-like
approximation with almost constant $\epsilon $ in between the threshold and some
higher energy is not excluded. Unfortunately, the data about $\rho (\omega )$
are obtained from so small transferred momenta that at LHC we hardly get any.
Experimentally, different possibilities can be verified by measuring the 
transverse momenta and energy spectra of particles in the humps.

With increments of the increase of the process (\ref{d1}) known, it is possible
to estimate the corresponding effective time in (\ref{green}). Using
$\Delta \epsilon _{eff}\approx 10^{-5}$ one gets it about $10^4$ times larger 
than the typical hadronic scale. This implies that usual hadronization effects
proceed much earlier than the full instability develops, i.e. a characteristic
time of instability growth in the high energy region is much larger than other 
time scales of the parton system evolution. It is known \cite{mr1, mr2} that
instabilities at lower frequencies may play more important role. Their
experimental verification is still awaited.

The roots $k_m(\omega )$ of Eqs. (\ref{plas}), (\ref{bunch}) or the poles
in Eq. (\ref{green}) are found at
${\rm Im}(\omega )\rightarrow +\infty $ both in upper and lower
halves of the $k$-plane. According to Starrock criteria \cite{kruk} it implies 
the absolute instability both for proper oscillations and Cherenkov radiation.

\section{Conclusion}

It is argued that the dispersion law of the quark-gluon medium at high
energies differs from that in the electromagnetism. Its main feature is
the excess of the chromopermittivity over 1. The definite model is considered.
The dispersion equation of in-medium gluodynamics is solved.
It is shown that the quark-gluon medium is unstable and responses to
external high energy partons by creation of Cherenkov gluons with specific
properties. 
\bigskip

{\bf Acknowledgments }

I thank A.A. Rukhadze and A.V. Leonidov for useful discussions. 

This work was supported by RFBR grants 09-02-00741; 08-02-91000-CERN and
by RAN programs.



\begin{thebibliography}{99}
\bibitem{mrow}
Mrowczynski S and Thoma M H 2007 {\it Ann Rev Nucl Part Sci} {\bf 57} 61
\bibitem{psmi}
Peigne S and Smilga A V 2009 {\it Uspekhi Fiz Nauk} {\bf 179} 697
\bibitem{inmed}
Dremin I M 2008 {\it Eur Phys J} C {\bf 56} 81
\bibitem{dpri}
Djongolov M K, Pisov S and Rizov V 2004 {\it J Phys} G {\bf 30} 425
\bibitem{okun}
Dremin I M 2010 {\it Yadernaya Fizika} {\bf 73}(4) 1823; 2010 {\it Phys Atom 
Nucl}; arXiv:0903.2941
\bibitem{kruk}
Kuzelev M V and Rukhadze A A 2007 {\it Metody teorii voln v sredah s dispersiey}
M Fizmatlit; 2009 {\it Methods of wave theory in dispersive media} 
Singapore WSPC
\bibitem{apan} 
Apanasenko A V {\it et al} 1979 {\it JETP Lett} {\bf 30} 145 
\bibitem{d1} 
Dremin I M 1979 {\it JETP Lett} {\bf 30} 140
\bibitem{d0}
Dremin I M 1981 {\it Sov J Nucl Phys} {\bf 33} 726
\bibitem{gr}
Grichine V M 2002 {\it Nucl Instr Meth} A {\bf 482} 629
\bibitem{dklv}
Dremin I M, Kirakosyan M R, Leonidov A V and Vinogradov A V 2009 
{\it Nucl Phys} A {\bf 826} 190
\bibitem{ja}
Jackson J D 1998 {\it Classical electrodynamics} John Wiley and Sons Inc 
 Fig 7.9
\bibitem{fe}
Feynman R P, Leighton R B and Sands M 1963 {\it The Feynman Lectures in Physics}
Addison-Wesley PC Inc vol~1, ch~31
\bibitem{kk}
Kalashnikov O K and Klimov V V 1980 {\it Sov J Nucl Phys} {\bf 31} 699
\bibitem{kl}
Klimov V V 1981 {\it Sov J Nucl Phys} {\bf 33} 934; 
1982 {\it Sov Phys JETP} {\bf 55} 199
\bibitem{we}
Weldon H A 1982 {\it Phys Rev} D {\bf 26} 1394
\bibitem{bi}
Blaizot J P and Iancu E 2002 {\it Phys Rep} {\bf 359} 355
\bibitem{rrs}
Rebhan A, Romantschke P and Strickland M 2005 {\it Phys Rev Lett} {\bf 94}
102303; 2005 {\it JHEP} {\bf 0509} 041
\bibitem{amy}
Arnold P, Moore G D and Yaffe L G 2005 {\it 72} 054003
\bibitem{ko} 
Koch V, Majumder A and Wang X N 2006 {\it Phys Rev Lett} {\bf 96} 172302
\bibitem{go}
Goldberger M and Watson K 1964 {\it Collision Theory} John Wiley and Sons Inc
Ch 11, sect~3, sect~4
\bibitem{scad}
Scadron M D 1979 {\it Advanced quantum theory and its applications through 
Feynman diagrams} Springer-Verlag p 326
\bibitem{da}
Damjanovic S {\it et al} (NA60) 2006 {\it Phys Rev Lett} {\bf 96} 162302
\bibitem{dnec}
Dremin I M and Nechitailo V A 2009 {\it Int J Mod Phys} A {\bf 24} 1221
\bibitem{drem1} 
Dremin I M 2006 {\it Nucl Phys} A {\bf 767} 233
\bibitem{4}
Trnka D {\it et al} 2005 {\it Phys Rev Lett} {\bf 94} 192303  
\bibitem{5}
Naruki M {\it et al} (KEK) 2006 {\it Phys Rev Lett} {\bf 96} 092301
\bibitem{Muto}  
Muto R {\it et al} (KEK) 2007 {\it Phys Rev Lett} {\bf 98} 042501 
\bibitem{6}
Kozlov A (PHENIX) 2006 arXiv:nucl-ex/0611025
\bibitem{7}
Kotulla M (CBELSA/TAPS) 2006 arXiv:nucl-ex/0609012
\bibitem{fw}
Wang F (STAR) 2004 {\it J Phys} G {\bf 30} S1299
\bibitem{ph} 
Adare A {\it et al} (PHENIX) 2007 arXiv:0705.3238; 0801.4545
\bibitem{ul} 
Ulery J G 2007 arXiv:0709.1633
\bibitem{ajit}                       
Ajitanand N N (PHENIX) 2007 {\it Nucl Phys} A {\bf 783} 519; arXiv:nucl-ex/0609038.
\bibitem{abel}
Abelev B I {\it et al} 2009 {\it Phys Rev Lett} {\bf 102} 052302
\bibitem{jel}
Jelley J V 1958 {\it Cherenkov radiation and its applications} Pergamon Press 
p 13
\bibitem{wake}
Dremin I M 2010 {\it Mod Phys Lett} A ; arXiv:0911.3233
\bibitem{holz}
Holzmann W G 2009 arXiv:0907.4833 [nucl-ex]
\bibitem{mus1}
Chakraborty P, Mustafa M G and Thoma M H 2006 {\it Phys Rev} D {\bf 74} 094002
\bibitem{mus2}
Chakraborty P, Mustafa M G, Ray R and Thoma M H 2007 {\it J Phys} G {\bf 34} 214
\bibitem{rmu}
Ruppert J and Mueller B 2005 {\it Phys Lett} B {\bf 618} 123
\bibitem{betz}
Betz B {\it et al} 2009 {\it Phys Rev} C {\bf 79} 034902
\bibitem{ryaz}
Ryazanov M I 1984 {\it Elektrodynamika condensirovannyh sred} M Nauka p 212
\bibitem{rukh}
Rukhadze A A 1962 {\it ZhTF} {\bf 62} 669 
\bibitem{dnaz}
Dremin I M and Nazirov M T 1983 {\it Pisma v ZHETF} {\bf 37} 198
\bibitem{ryzh}
Ryzhii V I 1968 {\it Pisma v ZHETF} {\bf 7} 28
\bibitem{eles}
Elesin V F 2004 {\it Uspekhi Fiz Nauk} {\bf 175} 197
\bibitem{aruk}
Alexandrov A F and Rukhadze A A 2002 {\it Lekzii po electrodynamike plasmopodobnyh
sred; neravnovesnye sredy} M Fizicheskii facultet MGU p 112
\bibitem{sruk}
Silin V P and Rukhadze A A 1961 {\it Electromagnitnye svoystva plasmy i 
plasmopodobnyh sred} M Gosatomizdat p 173
\bibitem{ginz}
Ginzburg V L 1967 {\it Rasprostranenie electromagnitnyh voln v plasme} M Nauka;
1970 {\it The propagation of electromagnetic waves in plasmas} Oxford New York
Pergamon Press
\bibitem{mr1}
Mrowczynski S 1988 {\it Phys Lett} B {\bf 214} 587
\bibitem{mr2}
Mrowczynski S 1993 {\it Phys Lett} B {\bf 314} 2197

\end{thebibliography}
\end{document}